\documentclass[aps,prl,twocolumn,showpacs]{revtex4}

\usepackage{amsmath,amssymb,latexsym}
\usepackage{graphicx}
\usepackage{amsfonts}
\usepackage{amsthm}
\usepackage{stmaryrd}
\usepackage{color}

\newcommand{\id}{\textrm{d}}

\begin{document}

\title{Monotone return to steady nonequilibrium}
\author{Christian Maes}
\affiliation{Instituut voor Theoretische Fysica, K.U.Leuven,
Belgium}
\author{Karel Neto\v{c}n\'{y}}
\affiliation{Institute of Physics AS CR, Prague, Czech Republic}
\author{Bram Wynants}
\affiliation{Institut de Physique Th\'eorique, CEA-Saclay, Paris, France}

\begin{abstract}
We propose and analyze a new candidate Lyapunov function for relaxation towards general nonequilibrium steady states. The proposed functional is obtained from the large time asymptotics of time-symmetric fluctuations. For driven Markov jump or diffusion processes it measures an excess in dynamical activity rates. We present numerical evidence and we report on a rigorous argument for its monotonous time-dependence close to the steady nonequilibrium or in general after a long enough time. This is in contrast with the behavior of approximate Lyapunov functions based on entropy production that when driven far from equilibrium often keep exhibiting temporal oscillations even close to stationarity.
\end{abstract}
\pacs{
05.70.Ln,  
05.40.-a   
}
\maketitle


Steady nonequilibrium refers to a time-homogeneous regime in which currents are maintained through an open system.  These currents displace particles, charges or energy between reservoirs that together constitute the environment of the system.  The different reservoirs are conveniently fixed as spatially separated and macroscopic, with fast relaxation regarding the coupling to the system, and each being kept at a thermodynamic equilibrium for specific temperature, pressure and chemical potential.  The phenomenology ranges from small systems like molecular motors producing mechanical work by hydrolysis of ATP \cite{bus}, to large-scale systems like the earth atmosphere exhibiting energy flows between the poles and the equator \cite{pal}.  On the relevant time-scales the stability of these
manifestly nonequilibrium systems  to small disturbances follows from their mere existence, but has not been described theoretically in a unifying
framework. Previously, most efforts concentrated on the linear irreversible regime and on the entropy production in the context of irreversible thermodynamics, \cite{gp}. In contrast, our work goes beyond the linear irreversible regime, fits within recent efforts in the construction of nonequilibrium statistical mechanics and, most importantly, is \emph{not} formulated in terms of entropy and its changes.

When the system is weakly coupled to the ``frustrating'' environment we obtain Markov processes effectively describing the stochastic dynamics of the relevant degrees of freedom,~\cite{vanhove}, and nonequilibrium is associated to the violation of detailed balance. In these circumstances the stability of the steady regime is mathematically described by identifying a large basin of initial conditions that are all
dynamically attracted to a unique steady state with some well defined stationary distribution. In equilibrium however, when the system is passive and no net currents are flowing, a much stronger and more relevant notion of stability is available: the Helmholtz-Gibbs variational principle characterizes equilibrium
as extremizing thermodynamic potentials such as entropy, enthalpy or free energies depending on the constraints.
At the same time these (equilibrium) thermodynamic potentials are related to heat and work and their derivatives inform us about the response to external stimuli.  The relaxation to equilibrium is accompanied by a loss of the system free energy, or an increase of entropy in the universe.  H-theorems, first formulated by Boltzmann for a dilute gas, add explicit witnesses, sometimes called H-functionals or Lyapunov functions, that show monotone behavior in time towards their equilibrium value.

So far, little on a same level of generality has been reported for the relaxation to steady nonequilibria, while all the same their stability is plainly visible already in daily observations. The present letter addresses this issue of monotone return for nonequilibrium systems as modeled via Markov processes that are traditionally used in the effective description on mesoscopic scales. Our approach develops a similar framework as works well in equilibrium, yet it differs in one important aspect: instead of starting from \emph{static} fluctuation theory which gives the relative entropy as Lyapunov function, we consider here its \emph{dynamical} counterpart.  The dynamical activity is a kinetic measure of noise or reactivity through which the system changes its state in the course of time. It has been found relevant for glassy dynamics and in kinetically constrained models, \cite{chan}, and it enables a more precise formulation of some older ideas due to Landauer \cite{land}.  In this letter we show that a particular excess dynamical activity yields a nonequilibrium Lyapunov function. The resulting variational principle and the related H-theorem also provide a natural generalization and correction to existing but approximative entropy production principles.\\

\emph{Monotonicity from static fluctuations.}
Before going into the details of our proposal, we first recall the equilibrium case by which we mean that detailed balance is satisfied. Let us stick to descriptions in terms of Markov processes say for a multilevel system in contact with a single thermal bath at inverse temperature $\beta$.  The corresponding Markov evolution is then given in terms of transition rates $k_0(x,y)$ between states (levels) $x\rightarrow y$ satisfying the condition of detailed balance $k_0(x,y)/k_0(y,x) = \exp \beta [U(x) - U(y)]$ with respect to energy function $U$.  The stationary distribution is given by the Boltzmann-Gibbs factor $\rho_0(x) = \exp [-\beta U(x)] /  Z$.  The evolution of a probability distribution $\mu_t$ satisfies the Master equation $\dot{\mu}_t(x) = \sum_y [\mu_t(y)k_0(y,x) - \mu_t(x)k_0(x,y)]$ for which, under standard conditions such as irreducibility,
$\mu_t\rightarrow \rho_0$ exponentially fast.  Much more is true however.  The free energy functional
$F(\mu) \equiv \sum_x \mu(x) [U(x) + \beta^{-1} \ln \mu(x)]$ {\it decreases} in time to its equilibrium value, $F(\mu_t) \rightarrow F(\rho_0) = -\beta^{-1}\ln Z$.
The mathematical reason is that the relative entropy
\begin{equation}\label{monot}
  s(\mu_t \,|\, \rho) = \sum_x \mu_t(x) \,\ln \frac{\mu_t(x)}{\rho(x)}
\end{equation}
of the evolved distribution with respect to the presumed unique stationary distribution $\rho$ (even when totally unknown) is {\it always} decreasing to zero,
while {\it for equilibrium},
$s(\mu_t \,|\, \rho_0) = \beta [F(\mu_t) - F(\rho_0)]$. Hence,
for a detailed balance dynamics we can take the free energy to serve as (equilibrium) Lyapunov function.

While the monotonicity of \eqref{monot} follows from a general convexity argument, there is a deeper and physically more relevant reason for it at least for equilibrium systems.  There the free energy also appears as fluctuation functional, much in the spirit of the
Boltzmann--Planck formula $S = k_B \log W$ for the thermodynamic entropy $S$ in terms of the ``plausibility'' $W$ of a macroscopic condition.  Indeed, as emphasized and heavily employed already by Einstein and later Onsager the formula $\exp (S/k_B)$ can be used to estimate the relative occurrence of equilibrium fluctuations.  Similarly, the free energy $F(\mu)$ governs the fluctuations of an ensemble of $N$ independent copies of the equilibrium system, in the sense that asymptotically for $N\rightarrow \infty$,
\begin{equation}\label{sta}
  \text{Prob}\Bigl[ \frac 1{N}\sum_{i=1}^N \delta_{x_i, x}
  \simeq \mu(x), \forall x \Bigr] = e^{- \beta N\,[F(\mu)-F(\rho_0)]}
\end{equation}
In other words, the ensemble frequency of state occupations is given by $\mu$ with an exponentially small probability when $\mu\neq \rho_0$,
while, by the law of large numbers, the probability in \eqref{sta} approaches one for $\mu=\rho_0$. That also explains why the Gibbs variational principle holds true.

In the absence of detailed balance the free energy difference in (\ref{sta}) should be replaced by the relative entropy (\ref{monot}).
 The relative entropy can therefore be taken as a static fluctuation potential and starting point to construct (generalized) thermodynamic potentials ~\cite{jona}. However, only for detailed balanced dynamics corresponding to equilibrium, has \eqref{monot} the usual interpretation in terms of thermodynamic potentials. Moreover and much more important, away from equilibrium fundamentally new observables related to the kinetics and to dynamical fluctuations become relevant and it is hence desirable to look for a natural {\it dynamic} alternative to the concept of relative entropy. We give such a proposal in the sequel.\\

\emph{Monotonicity from dynamical fluctuations.}
In the case of dynamical fluctuations, one investigates deviations from typical time-averages.  Again we work with a Markov dynamics as above where we now have quite general jump rates $k(x,y)$ for an irreducible process with stationary distribution $\rho$.
Denoting by $x_t$ the state of the system at time $t$, we consider
\begin{equation}
p_{\tau}(x) = \frac 1{\tau}\int_0^\tau \delta_{x_t,x} \,\id t
\end{equation}
i.e., the fraction of time spent in each state $x$.  Also in this case a fluctuation formula holds, this time for large $\tau$,
\begin{equation}
  \text{Prob}\,\Bigl[p_\tau(x) \simeq \mu(x),\,\forall x\Bigr] = e^{-\tau D(\mu)}
\end{equation}
This $D(\mu)$ is the rate function for dynamical fluctuations. We call it the Dynamical Activity (DA).
It gives a dynamical analogue of \eqref{sta}, see also~\cite{jac}.  Like for static fluctuations we have the immediate property that $D(\mu)\geq D(\rho) = 0$, expressing the fact that the probability of a fluctuation $\mu$ tends to zero for large times and $p_{\tau}(x)\to\rho(x)$. The central question in this paper however asks more: does  $D(\mu_t)$ decrease monotonically in time when $\mu_t$ evolves according to the Master equation? For this we need to see first how we can compute this DA.
 Following the results of \cite{mprf}, the following algorithm can be given.
First, we modify the transition rates by applying an extra potential $A$,
\begin{equation}\label{inv1}
  k(x,y) \rightarrow
  k_A(x,y) = k(x,y)\exp \Bigl[ \frac{A(y)-A(x)}{2} \Bigr]
\end{equation}
to define
\[
\tilde{D}(\mu,A) = \sum_{x,y} \mu(x) \bigl[ k(x,y) - k_A(x,y) \bigl]
\]
Note that $\sum_y k(x,y)$ are the escape rates. So $\tilde{D}(\mu,A)$ is given in terms of an \emph{excess} in expected escape rate when mutually comparing the original with the $A-$modified system, when occupations are distributed with $\mu$.
The DA is then obtained from the following equations:
\begin{eqnarray}
&& D(\mu) = \tilde{D}(\mu,V)\nonumber \\
&&\sum_y \bigl[ k_V(x,y)\,\mu(x) - k_V(y,x)\,\mu(y) \bigr] = 0 \label{defV}
\end{eqnarray}
In words, given a probability distribution $\mu$ we find the potential $V$, or the extra force $V(x) - V(y)$ applied over the transition $x\rightarrow y$ such that now $\mu$ becomes stationary for the modified system with transition rates $k_V(x,y)$.
In fact, we prove in \cite{math} that this problem has a unique (up to a constant) solution. The construction above makes contact with the blowtorch theorem of Landauer, \cite{land}, since the modification of the dynamics can be interpreted as inserting an extra possibility of energy exchange.  The extra potential $V$ can in that same sense be called the {\it blowtorch} and $D(\mu)$ monitors changes in the kinetics.

An advantage over working with \eqref{monot}, in principle, is that the definition of the DA does not depend on the stationary distribution.
Yet, finding the correct $V$ for a given distribution $\mu$ entails solving the inverse problem \eqref{defV}, and is therefore
analytically at least as challenging as finding the stationary distribution (needed for static fluctuations).
However, a straightforward computation shows that the function $\tilde{D}(\mu,A)$ is \emph{concave} as a function of $A$ and is \emph{maximal} at $\tilde{D}(\mu,V) = D(\mu)$. Therefore, for computing the DA one can use all the numerical power of maximizing algorithms.\\

\emph{Relation to entropy production principles.}
The instantaneous Entropy Production Rate (EPR), \cite{schnak}, is given by
\begin{equation}
  {\cal{E}}(\mu)=\sum_{x,y}\mu(x)k(x,y)
   \ln\frac{\mu(x)k(x,y)}{\mu(y)k(y,x)}
\end{equation}
Only under detailed balance does EPR coincide with minus the time-derivative of \eqref{monot}.
In~\cite{stijn} minimum and maximum entropy production principles for Markov processes as in e.g.~\cite{klein} were understood in terms of dynamical fluctuation theory. The key-point is a remarkable simplification of the above formula for $D(\mu)$ in the close-to-equilibrium regime where detailed balance is only weakly broken, and for $\mu$ not too different from the stationary distribution $\rho$. Under these special conditions one finds
that $D(\mu) \simeq [{\cal{E}}(\mu) - {\cal{E}}(\rho)]/4$.
It is exactly this relation to the dynamical fluctuation law that lies behind the close-to-equilibrium validity of various entropy production principles. In that same regime, the EPR is also a Lyapunov functional.
Although this EPR continues to make sense as dissipation rate far from equilibrium, it is then no longer related to the DA, nor does it give  a minimum entropy production principle, nor is it monotonous in time.\\

\emph{$D(\mu)$ as a Lyapunov functional.}
We present here the evidence that our dynamical activity functional $D(\mu_t)$ becomes monotonically decreasing to its stationary nonequilibrium value $D(\rho) = 0$ substantially beyond the close-to-equilibrium regime, in contrast to the EPR.\\
We first demonstrate our hypothesis numerically via examples for which the DA has been computed by the variational algorithm indicated above.   As our first example we consider
the simple symmetric exclusion process on a lattice interval with open boundaries where different particle densities are imposed left and right.
It constitutes a paradigmatic nonequilibrium model that for example simulates effects of ion hopping through small channels.
Because the driving is via the open boundary, the system is further away from equilibrium for smaller sizes --- to be specific, we have taken a sequence of 6 sites.
For this system we have numerically checked the time-evolution of both the fluctuation functional $D(\mu_t)$
and ${\cal{E}}(\mu_t)$. Several initial distributions have been explored.
We find that for a large range of different particle densities imposed at the open ends and for different initial distributions, both functionals monotonically decrease in time.
One might be tempted to conclude from this that the EPR is as good a candidate to be a Lyapunov function as $D(\mu)$. However, this is not true, as the next example demonstrates.

\begin{figure}[ht]
\includegraphics[width=0.50\linewidth]{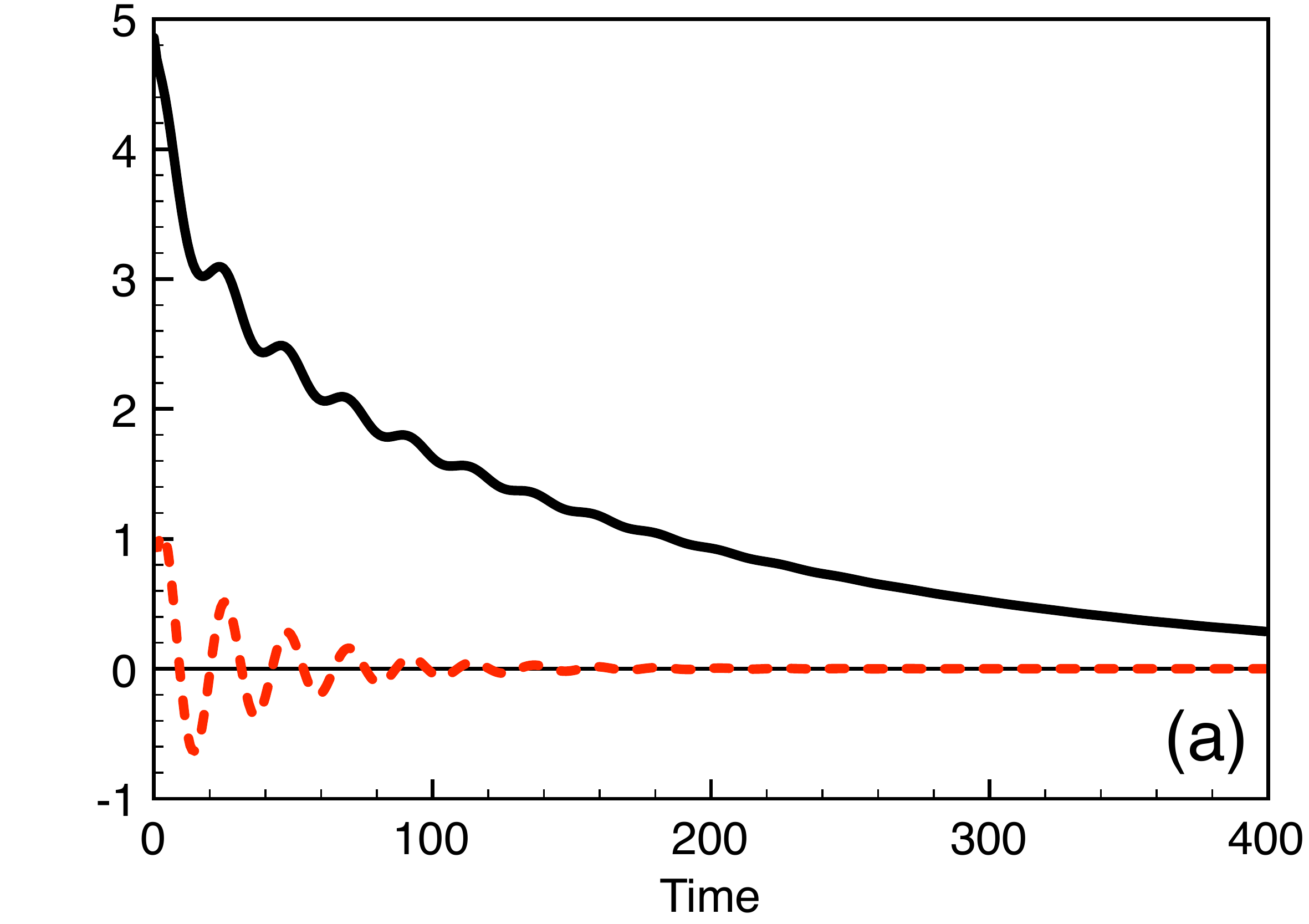}
\includegraphics[width=0.48\linewidth]{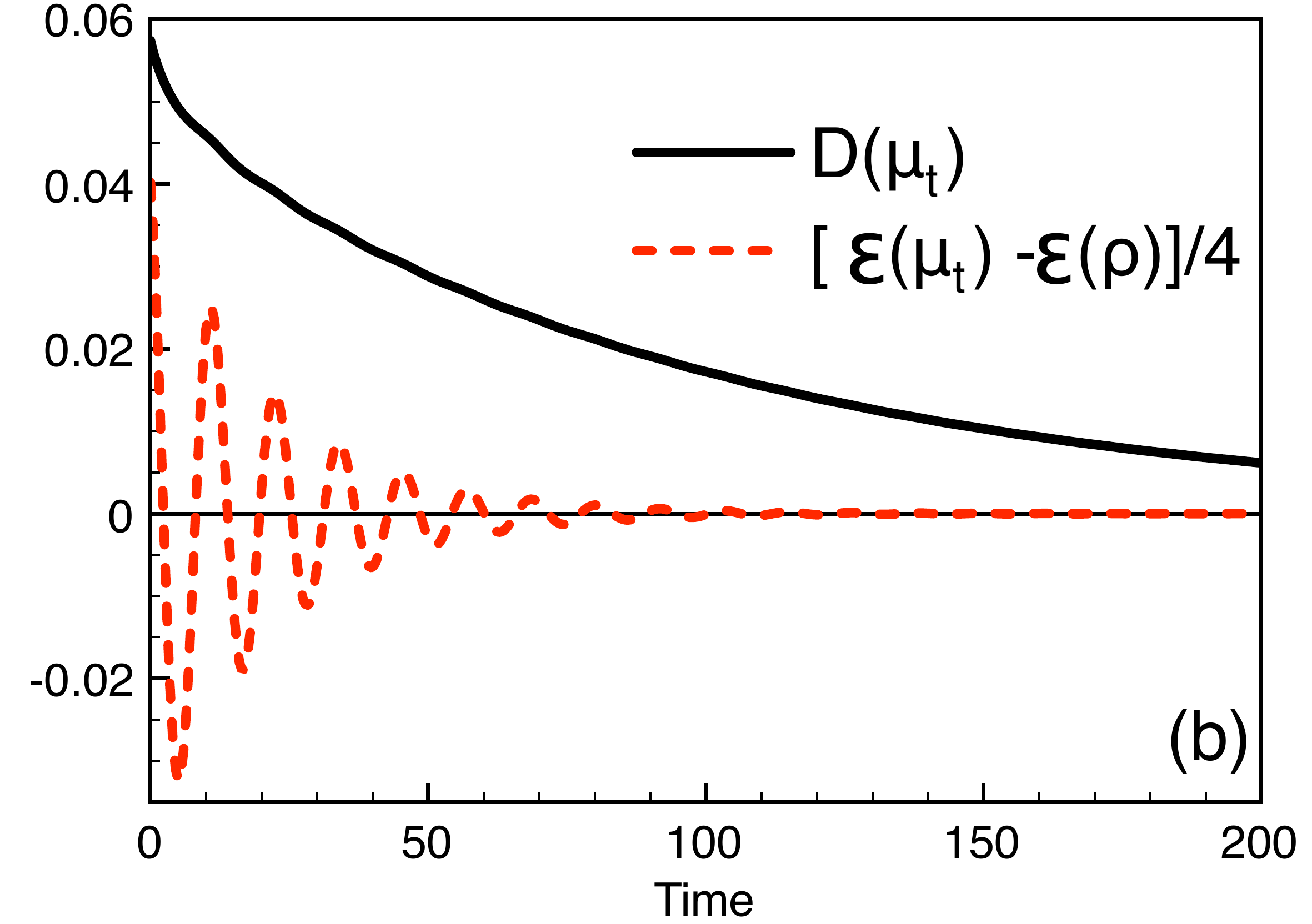}
\caption{The functionals $D(\mu_t)$ and $[{\cal{E}}(\mu_t) - {\cal{E}}(\rho)]/4$ for driven diffusion, with $N=100,  \beta =1$, $f=5$ and $U(x) = \cos(6\pi x)$. In the case (a) the initial condition is far from stationarity: $\mu_0(x) = 1/2 + 1/2N$ for $x=0$ and $\mu_0=1/2N$ otherwise. In (b) $\mu_0(x) = \rho(x) + d\, \nu(x)$  with $d = 0.1$,
$\nu(x) = \pm 1/N$ for $x\lessgtr 0.5$.}
\label{fig:ring}
\end{figure}

As our second example we have checked driven diffusion on a ring, which corresponds to the motion of a bulk-driven particle in a toroidal trap. It is experimentally very accessible, including its linear response behavior far-from-equilibrium~\cite{lyo}.  We model this by a Markov jump process on a ring with $N$ sites. The particle can hop one site to the left or right with hopping rates
\begin{equation}
  k\bigl(x,x\pm \frac{1}{N}\bigr) =
  \exp\bigl\{ -\beta\bigl[ U\bigl(x\pm \frac{1}{N}\bigr) - U(x) \mp \frac{f}{N} \bigr] \bigr\}
\end{equation}
which can be seen as a space-discretized version of the overdamped diffusion dynamics driven by the force $f$,
\begin{equation}
  \dot{x}_t = f - \frac{\id U}{\id x}(x_t) + \sqrt{\frac{2}{\beta}}\,\zeta_t
\end{equation}
with $\zeta_t$ standard white noise.  Again we have numerically calculated both $D(\mu_t)$ and ${\cal{E}}(\mu_t)$  with various initial distributions and each time with forcing $f$ comparable in size to $dU/dx$. The main difference between the two functions ${\cal{E}}(\mu_t)$ and $D(\mu_t)$ is illustrated in Fig.~\ref{fig:ring}: in (a) the initial condition is far from stationarity. We see that $D(\mu_t)$ stops oscillating after some short time and becomes monotone, while the entropy production keeps oscillating all the way. This is made more clear in (b), where the initial distribution was taken close to the nonequilibrium stationary distribution.

We finally turn to analytical arguments and we employ a strategy that is common in linear stability theory: we linearize around the stationary distribution $\rho$ which amounts to keeping only the leading order in the deviation $\mu - \rho$. That is exactly what we need for the long time behavior of the nonequilibrium system where indeed $\mu_t$ gets closer to the stationary $\rho$ as time $t$ runs. In this approximation we have found
\begin{equation}\label{DActs}
 \frac{\id}{\id t}D(\mu_t) =  -\frac{1}{2}\,\big[\left<(LV_t)^2\right>_{\rho}+\left<V_tL^2V_t\right>_{\rho}\big]
\end{equation}
in terms of the stationary expectations $\langle\cdot\rangle_\rho$ and with $L$ the linear operator on state functions defined by $(LV)(x)  = \sum_yk(x,y)[V(y)-V(x)]$ (called, backward Markov generator).
From (\ref{DActs}) conditions for the monotonicity can be formulated.
E.g. $(\ref{DActs})$ immediately implies the following sufficient condition: $\left<fL^2f\right>\geq 0$ $\forall f$, which can be reformulated in terms of the spectrum of $L$, see \cite{math}. We want to focus however on necessary and sufficient conditions
 in terms of linear response behavior. This nonequilibrium response theory was e.g.\ studied in \cite{prl}: the transition rates of the system are modified as in (\ref{inv1}) with $A = h_{\tau}V$ (depending on time $\tau$), so that an expression can be obtained for the
auto-response function $R_{V}(\tau) = \frac{\delta}{\delta h_0}\langle V(x_{\tau})\rangle\big|_{h=0}$ when perturbing with potential $V$ at time zero starting from stationarity.  Remarkably, that exactly gives the right-hand side of \eqref{DActs}, or,
\begin{equation}
 \left.\frac{\id}{\id \tau} R_{V_t}(\tau)\right|_{\tau=0} = \frac{\id}{\id t}D(\mu_{t})
\end{equation}
where $V_t$ remains the potential as in \eqref{defV} determined by the distribution $\mu_{t}$.
We therefore see that the long time monotonicity of $D(\mu_t)$ is directly equivalent with ``normal'' behavior of the auto-response. By ``normal'' we mean that the auto-response always starts out decaying in time.
We leave the mathematical details to~\cite{math}.

\emph{Conclusion and outlook.}
The dynamical activity functional~\eqref{defV} is proposed as a new Lyapunov functional for the long time behavior of Markov evolutions.   We have collected numerical evidence for some standard nonequilibrium models.  Mathematically, we have found the equivalence between ``normal''  linear auto-response at short times and monotonicity of DA at long times. The rigorous formulation and proofs are collected in~\cite{math}, supporting and explaining our numerical findings.

An important motivation to search for Lyapunov functions in the relaxation to nonequilibrium steady states comes from the problem of stability of competing thermodynamic regimes in complex driven systems. Towards this application, the behavior of the dynamical activity needs to be further analyzed in the case of systems with a larger number of degrees of freedom. In order to recognize and exploit the dynamical activity and its changes as a thermodynamic state function, a more operational definition in terms of suitable thermodynamic processes is needed. This appears now to be one of the major relevant open problems when attempting to build a nonequilibrium thermodynamic formalism beyond the by now well understood linear irreversible regime.  A second interesting application relates to the notion of statistical forces outside equilibrium.  Maximum entropy arguments govern the relaxation to classical and quantum equilibrium alike.  Outside equilibrium, from our present investigations, it appears that the minimization of dynamical activity \eqref{defV} (called {\it motion out of noisy states} in \cite{land}) might non-trivially characterize the steady behavior.

\paragraph*{Acknowledgments:}
K.N.\ acknowledges the
support from the Academy of Sciences of the Czech Republic under Project No.~AV0Z10100520.

\bibliographystyle{plain}

\begin{thebibliography}{10}

\bibitem{bus}
C. Bustamante, Y.R. Chemla, N.R. Forde and D. Izhaky,
\emph{Annu. Rev. Biochem.} {\bf 73}, 705–-748 (2004).

\bibitem{pal}
G. W. Paltridge, \emph{Nature} {\bf 279}, 630 (1979).

\bibitem{gp}
P.~Glansdorff and I.~Prigogine,
\textsl{Thermodynamic Theory of Structure, Stability, and Fluctuations},
Wiley, London (1971);
J.~Keizer, \textsl{Statistical Thermodynamics of Nonequilibrium Processes}, Springer-Verlag, New York (1987);
T.~Wilhelm and P.~H\"anggi,
\emph{J. Chem. Phys.} {\bf 110}, 6128 (1999).


\bibitem{vanhove}
L.~A.~B\'anyai, \textsl{Lecures on non-equilibrium theory of condensed matter}, pp.~149--158, World Scientific Publishing Co (2006);
J.~L.~Lebowitz and H.~Spohn,
\emph{Adv. Chem. Phys.} {\bf 38}, 109 (1978).


\bibitem{chan}
R.~Jack, J.P.~Garrahan, D.~Chandler,
\emph{J.Chem.Phys.} {\bf 125}, 184509, (2006);
J.~P.~Garrahan, R.L.~Jack, V.~Lecomte, E.~Pitard, K.~van
Duijvendijk, and F.~van Wijland,
{\it J.~Phys. A: Math. Gen.}  {\bf 42}, 075007 (2009).

\bibitem{land}
R.~Landauer,
{\it J. Stat. Phys.} {\bf 53}, 233--248 (1988).


\bibitem{jona}
L.~Bertini, A.~De Sole, D.~Gabrielli, G.~Jona-Lasinio, C.~Landim, {\it Phys. Rev. Lett.} {\bf 87}, 040601 (2001); {\it J. Stat. Phys.} {\bf 107}, 635 (2002).


\bibitem{math}
C.~Maes, K.~Neto\v{c}n\'y, and B.~Wynants,
{\tt arXiv:1102.2690}.



\bibitem{jac}
R.~L.~Jack and P.~Sollich,
\emph{Prog. Theor. Phys.} Supplement No. 184, pp. 304--317 (2010).

\bibitem{mprf}
C.~Maes, K.~Neto\v{c}n\'y, and B.~Wynants,
\emph{Markov Proc. Rel. Fields.} {\bf 14}, 445--464 (2008);
C.~Maes, K.~Neto\v{c}n\'y, and B.~Wynants, \emph{Physica A} {\bf 387}, 2675--2689 (2008)

\bibitem{stijn}
C.~Maes and K.~Neto\v{c}n\'y,
\emph{J. Math. Phys.}
\textbf{\bf 48}, 053306 (2007).

\bibitem{klein}
M.~J.~Klein and P.~H.~E.~Meijer,
\emph{Phys.~Rev.} \textbf{96}, 250 (1954);
Luo Jiu-li, C.~Van den Broeck, and G.~Nicolis,
\emph{Z. Phys. B Condensed Matter} {\bf 56}, 165 (1984).

\bibitem{schnak}
J.~Schnakenberg, \emph{Rev. Mod. Phys.} \textbf{48}, 571 (1976);
D.~J.~Evans, D.~J.~Searles and S.~R.~Williams, \emph{J. Stat. Mech.} {\bf 7}, P07029 (2009).

\bibitem{lyo}
Juan Ruben Gomez-Solano, A.~Petrosyan, S.~Ciliberto and C.~Maes, {\it J.Stat. Mech.}, P01008 (2011).


\bibitem{prl}
M.~Baiesi, C.~Maes and B.~Wynants, \emph{Phys. Rev. Lett.} {\bf 103}, 010602 (2009).

\end{thebibliography}

\end{document}